\documentclass[twocolumn,showpacs,preprintnumbers,amsmath,amssymb,showkeys]{revtex4}

\usepackage{amsmath,amsfonts,amsbsy}
\usepackage{dsfont}

\usepackage{xy}
\xyoption{matrix} \xyoption{arrow} \xyoption{arc} \xyoption{color}

\usepackage{subfigure}
\usepackage{tikz}
\usetikzlibrary{decorations.pathmorphing,shapes,arrows}
\usepackage{yhmath}

\usepackage{enumerate}
\usepackage{enumitem}
\makeatletter
\def\namedlabel#1#2{\begingroup
    #2%
    \def\@currentlabel{#2}%
    \phantomsection\label{#1}\endgroup
}
\makeatother

\usepackage[latin1]{inputenc}  

\newcommand{\Id}{\mathds{1}}
\newcommand{\B}{\mathbb{B}}

\newcommand{\R}{\mathbb{R}}
\newcommand{\Z}{\mathbb{Z}}

\newcommand{\T}{\mathbb{T}}

\newcommand{\A}{\mathcal{A}}

\newcommand{\scal}[2]{\left\langle #1 | #2 \right\rangle}

\newcommand{\bra}[1]{\left\langle #1 \right|}
\newcommand{\ket}[1]{\left| #1 \right\rangle}
\newcommand{\eu}{\mathrm{e}}
\newcommand{\iu}{\mathrm{i}}
\newcommand{\di}{\mathrm{d}}

\newcommand{\bk}{\mathbf{k}}

\newcommand{\bx}{\mathbf{x}}
\newcommand{\bg}{\mathbf{g}}

\DeclareMathOperator{\Tr}{Tr}

\DeclareMathOperator{\dist}{dist}

\newcommand{\ie}{{\sl i.\,e.\ }}
\newcommand{\eg}{{\sl e.\,g.\ }}
\newcommand{\set}[1]{ \left\{  #1 \right\}}

\newcommand{\half}{\mbox{\footnotesize $\frac{1}{2}$}}

\newcommand{\crucial}[1]{{\it \textbf{#1}}}

\usepackage
{hyperref}

\begin{document}

\title{The Localization Dichotomy for gapped periodic quantum systems}

\author{Domenico Monaco$^{\star}$}
\author{Gianluca Panati$^{\diamond}$}
\email[Email: ]{panati@mat.uniroma.it}
\author{Adriano Pisante$^{\diamond}$}
\author{Stefan Teufel$^{\star}$}

\affiliation{$\star$: Eberhard Karls Universit\"at T\"ubingen, 72076 T\"ubingen, Germany}
\affiliation{$\diamond$:  ``Sapienza'' Universit\`{a} di Roma, 00185 Roma, Italy}

%
%

\date{December 30, 2016}

\begin{abstract} We investigate the localization properties of  gapped periodic quantum systems, modeled  
by a periodic or covariant family of projectors, as \eg the orthogonal projectors on the occupied orbitals 
at fixed crystal momentum for a gas of non-interacting electrons. We prove a general localization dichotomy for dimension $d \leq 3$: either the system is topologically trivial \ie all the Chern numbers vanish, or any arbitrary choice of composite 
Wannier functions yields an infinite expectation value for the squared position operator. Equivalently, in the topologically non-trivial phase, the localization functional introduced by Marzari and Vanderbilt diverges, as already noticed in the case of the Haldane model. Our result is formulated by using only the relevant symmetries of the system, and it is thus largely
model-independent.  Possible applications include both  tight-binding and continuous models of crystalline solids, cold gases in optical lattices as well as flat band superconductivity. 
\end{abstract}




\maketitle


The profound result by Thouless \textsl{et al.}\ \cite{TKNN82} first highlighted the relation between the transport 
properties of a gapped periodic quantum system and the topology of the space of  occupied states, 
decomposed with respect to crystal momentum.  
After a pioneering intuition by Haldane \cite{Haldane88}, this transport-topology relation appeared 
to be ubiquitous in condensed-matter  physics, being a key feature to understand 
 topological insulators and superconductors \cite{Kitaev09, HasanKane10, Ando13},     
cold atoms in optical lattices 
\cite{Trombettoni et al experimental 2005, Jotzu et al 2014, SmerziTrombettoni2003}, 
flat band superconductivity \cite{PeottaTorma2015} 
as well as macroscopic polarization and piezoelectric currents \cite{KSV, Resta94, PanatiSparberTeufel2009}.
Remarkably, topological transport is robust with respect to disorder and interactions,
as confirmed by recent rigorous results based on renormalization group techniques \cite{GMP16}: 
In the Hubbard-Haldane model, the transverse conductivity of a gas of interacting electrons is exactly equal 
to that of a non-interacting gas, provided the coupling constant is sufficiently small.


In this Letter, we show that, for any gapped periodic quantum system, the topology of the space of occupied states 
has profound consequences for 
the localization properties of the system.  
We are interested in periodic systems, where the eigenstates of the Hamiltonian (\eg Bloch orbitals for a non-interacting electron gas) are extended over  the whole space. Therefore, to distinguish between localized and delocalized states of the system, we use a finer notion of localization, which is provided by the maximally-localized composite Wannier functions (CWFs)
\cite{MaVa,Marzari_et_al12}.   
Specifically, we show that for every gapped periodic $d$-dimensional quantum system with $d \leq 3$
there is a  \crucial{localization dichotomy}, in the following sense:  \renewcommand{\labelenumi}{{\rm(\roman{enumi})}}
\begin{enumerate}[label=(\roman*)]
\vspace{-2mm}
\item \label{item:trivial} \textsl{either} the space of occupied states is topologically trivial, 
and correspondingly there exists a system of exponentially 
localized  composite Wannier functions,  
as shown in \cite{Wannier_letter_BPCM};
\vspace{-2mm}
\item \label{item:non-trivial} \textsl{or} any 
system of composite Wannier functions $w= (w_1,\ldots, w_m)$  yields 
a diverging second moment of the position operator, \ie
\begin{equation} \label{X^2explodes}
\langle X^2 \rangle_{w} =  \sum_{a=1}^m \int_{\R^d}  {|\bx|^2} \, |w_a(\bx)|^2 \di \bx  = + \infty. 
\end{equation} 
\end{enumerate} 
Notice that, whenever the system enjoys time-reversal  symmetry, either bosonic or fermionic, we are in case
 \ref{item:trivial}, as argued in \cite{Wannier_letter_BPCM,  Panati2007, MonacoPanati15}. Viceversa, breaking time-reversal symmetry yields generically to case \ref{item:non-trivial}.  Intermediate regimes are forbidden, as we prove below. 
We also prove that one can always choose a Bloch gauge such that the corresponding CWFs 
satisfy 
\begin{equation} \label{X^2s_control}
\sum_{a=1}^m \int_{\R^d}  |\bx|^{2s} \, |w_a(\bx)|^2 \di \bx  < + \infty  \quad  \forall s < 1.
\end{equation}
Assuming that the Wannier functions have a power-law decay at infinity as 
$w_a(\bx) \asymp |\bx|^{-r}$ for $|\bx| \rightarrow \infty$,  
the comparison between \eqref{X^2explodes} and \eqref{X^2s_control} yields 
the optimal asymptotic decay in the delocalized regime: 
 \begin{equation} \label{}
w_a(\bx) \asymp 
\begin{cases}
|\bx|^{-2}        & \text{ for } d=2,  \\
|\bx|^{-5/2}     & \text{ for } d=3. 
\end{cases}
\end{equation}
The $2$-dimensional result agrees with the prediction in \cite{Thouless1984} and with previous results
for the Haldane model \cite{ThonauserVanderbilt2006}.  {Our analysis is instead model-independent, 
it applies to both tight-binding and continuous models,  
as well as  to interacting electron gases within the Hartree-Fock approximation.} 
Notice that the previous asymptotic decay implies that, in the topologically non-trivial phase, 
the Marzari-Vanderbilt localization functional \cite{MaVa} diverges, as noticed in \cite{ThonauserVanderbilt2006}: 
we expect a logarithmic divergence to appear as the numerical mesh in $\bk$-space becomes finer and finer. 



The starting point of our analysis is a family of orthogonal projectors $\set{P(\bk)}$, 
where $P(\bk)$ projects on the manifold of occupied states at fixed crystal momentum $\bk$. 
For example, in the case of independent electrons in a crystal, one has
\begin{equation} \label{Fermi projector}
P(\bk) = \sum_{n=1}^m \ket{u_{n,\bk}} \bra{u_{n,\bk}}
\end{equation}
where $m$ is the number of Bloch bands below the Fermi energy (assumed to lie in an energy gap) and $u_{n,\bk}(\bx)$ is the periodic part of the Bloch function $\psi_{n,\bk}$: $\psi_{n,\bk}(\bx) = \eu^{\iu \bk \cdot \bx} u_{n,\bk}(\bx)$. It solves 
the eigenvalue equation for the operator
\begin{equation} \label{eqn:H(k)}
H(\bk) = \half \Big( -\iu \nabla_\bx - \tfrac{1}{c} A(\bx) + \bk \Big)^2 + V_{\rm per}(\bx)
\end{equation}
(in $\mathrm{Ha}$) where $V_{\rm per}$ is periodic with respect to a Bravais lattice, and $A(\bx)$ is either 
periodic (as in Chern insulators), or linear, \ie corresponding to a uniform magnetic field 
(as in Quantum Hall systems).  Similarly, one might also interpret  \eqref{Fermi projector}
as the orthogonal projector on the occupied Bloch orbitals of a tight-binding model, as the 
 Haldane \cite{Haldane88} or the Hofstadter model \cite{Hofstadter76}.   

In the case of interacting fermions, one considers instead the ground state of a periodic many-body 
Hamiltonian, as \eg the Hubbard-Haldane Hamiltonian \cite{GMP16}. In the thermodynamic limit, the 
corresponding $1$-body density matrix, here denoted by $\rho$, commutes with the lattice translations.  
Hence, it can be decomposed with respect to the crystal momentum as a (generalized) direct sum
\begin{equation} \label{rho decomposition}
\rho  =  \int^{\oplus}_{\T^d} \rho(\bk) \di\bk  
\end{equation}
where $\T^d$ is a $d$-dimensional torus.  
Within the Hartree-Fock (HF) approximation, both $\rho$ and $\rho(\bk)$ are orthogonal projectors
\cite{CancesDelaurenceLewin2008}, 
and the family $\set{P(\bk) \equiv \rho(\bk)}$  is thus the starting point of our analysis 
in the HF setting. Beyond the HF approximation, as \eg when highly-correlated electrons are considered,
our approach needs instead substantial modifications, which will be investigated in the future.     
In this broader setting the position operator is replaced by $\hat{X} = \sum_{i=1}^N \bx_i$ as in \cite{RestaSorella1999} and the Wannier transform should be appropriately modified.  



Since the system is gapped, one has that 
\begin{equation} \label{eqn:proj_smooth}
\bk \mapsto P(\bk)  \text{ is real analytic}  
\end{equation}
as shown in \cite{desCloizeaux64_A}, see \cite{Nenciu91, PanatiPisante}. 
Moreover, the family of projectors must be compatible with the 
periodicity of the system. This means that when $\bk$ is shifted by the reciprocal lattice vector $\bg$, one has  
\begin{equation} \label{eqn:tau-cov}
P(\bk+\bg) = \tau_{\bg} \, P(\bk) \, \tau_{\bg}^{-1} 
\end{equation}
where $ \tau_{\bg}$ are unitary operators such that  $\tau_{\bg_1} \tau_{\bg_2} =  \tau_{\bg_1 + \bg_2}$
for every $\bg_1,\bg_2$ in the reciprocal lattice $\mathcal{R}^*$. 
For example, in the case of non-interacting electrons $u_{n,\bk}(\bx)$ satisfies 
\begin{equation} \label{eqn:tau}
u_{n, \bk + \bg}(\bx) = \eu^{-\iu \, \bg \cdot \bx} \, u_{n,\bk}(\bx). 
\end{equation}
Therefore, the Fermi projector \eqref{Fermi projector} satisfies \eqref{eqn:tau-cov}, with $\tau_{\bg} = \eu^{-\iu \, \bg \cdot \bx}$.
For tight-binding models, one has instead $\tau_{\bg} = \Id$ for every $\bg$, \ie periodicity holds
\begin{equation} \label{eqn:P_periodic}
P(\bk+\bg) =  P(\bk)    \qquad \forall \bg \in \mathcal{R}^*. 
\end{equation}
For the sake of a concise exposition, in this letter we focus on the periodic case \eqref{eqn:P_periodic}, 
while the general case  \eqref{eqn:tau-cov} is detailed in the companion 
paper \cite{companion}.
 


It is by now well known that, in general, periodic 
orthonormal  frames $(u_{1,\bk}, \ldots, u_{m,\bk})$ consisting of Bloch functions which depend continuously on $\bk$ may not exist. The reason is twofold. First of all, a \crucial{local obstruction} appears at those point $\bk$ where two (or more) Bloch bands intersect each other. In one dimension, Kohn \cite{Kohn59} showed that it is possible to define Bloch states that are analytic functions of $\bk$. In two and three dimensions, this is generically not possible \cite{desCloizeaux64}. On the other hand, Blount \cite{Blount62} noticed that the 
regularity properties of Bloch functions can be improved by considering a set of states, corresponding to a \emph{composite energy band} \cite{desCloizeaux64}, which are separated by an energy gap from all others.  

We call \crucial{Bloch frame} an orthonormal system of functions  $(v_{1,\bk}, \ldots, v_{m,\bk})$  that span the same vector space as the $m$ Bloch states of the composite band, so that 
$ P(\bk) = \sum_{a} \ket{v_{a, \bk}}\bra{v_{a,\bk}}$.  In the HF many-body setting, the latter equation is instead 
the definition of a Bloch frame.
 
It can be shown that there always exists a choice of $v_{a,\bk}$ such that $\bk \mapsto v_{a,\bk}$ is locally continuous. However, patching these functions together in a continuous way can be achieved, in general, only at the price of losing the periodicity property \eqref{eqn:P_periodic} for the quasi-Bloch states. This is the origin of a second, \crucial{global obstruction}, namely the one to the existence of a globally continuous and periodic Bloch frame. The latter is of topological nature, and is encoded in the Chern number(s) provided $d \leq 3$ \cite{TKNN82, AvronSeilerSimon83, Simon83, Resta94}.  Whenever the system is   time-reversal symmetric, the latter global obstruction vanishes, and a  continuos (actually, analytic) periodic Bloch frame exists 
\cite{Wannier_letter_BPCM, Panati2007, MonacoPanati15}.
It can be explicitly constructed whenever the time-reversal operator  is even 
\cite{FiorenzaMonacoPanati16, CoHeNe2015, CLPS16}. 


Composite Wannier functions are  the counterpart, in position space, of quasi-Bloch functions. 
More precisely, given a periodic Bloch frame $v_{\bk} = (v_{1, \bk}, \ldots, v_{m,\bk})$ (not necessarily continuous), 
the corresponding system of composite Wannier functions $w = (w_1, \ldots, w_m)$ is defined by
\begin{equation} \label{Wannier}
w_a(\bx) : = \frac{1}{|\B|} \int_{\B}\eu^{ \iu \bk \cdot \bx} v_{a, \bk} (\bx)\, \di \bk 
\end{equation}
where $\B$ denotes the first (magnetic) Brillouin zone, and $|\B|$ its volume. 

Since the Wannier transform \eqref{Wannier} intertwines the operator $\iu \nabla_{\bk}$ with multiplication times $\bx$
\cite{Nenciu91, PanatiPisante}, the rate of decay of composite Wannier functions as $|\bx| \to \infty$ crucially depends on the regularity in $\bk$ of the corresponding Bloch frame. Quantitatively, a system of CWFs satisfies 
\begin{equation} \label{Finite X^2}
\sum_{a=1}^m \int_{\R^d}  {|\bx|^2} \, |w_a(\bx)|^2 \di \bx  < + \infty
\end{equation}
if and only if the corresponding Bloch frame satisfies 
\begin{equation} \label{eqn:Dirichlet energy}
\sum_{a=1}^m \int_{\B}  \di \bk   \int_{Y}  \di \bx \,\,  |\nabla_{\bk} v_{a, \bk} (\bx)|^2 < + \infty   
\end{equation}
where $Y$ is a centered unit cell for the Bravais lattice. 
More general relations can be obtained, relating the validity of the equation \eqref{X^2s_control}
to a constraint on the type of singularity the map $\bk \mapsto v_{\bk}$ may have, as detailed in \cite{companion}. 
For example, for $2$-dimensional systems equation  \eqref{X^2s_control} is 
compatible with a singularity like $v_{a,\bk} \asymp |\bk - \mathbf{k}_0|^{-1}$ for $\bk \to \mathbf{k}_0$, 
but not with a more severe one.    

\begin{figure}[ht]
\centering
\begin{tikzpicture}[scale=.32]
\draw [gray!70, dashed] (-2.165,-1.25) ++(-5,5) ++(4.33,2.5) -- ++(0,-10) -- ++(10,0) ++(-10,0) -- ++(-4.33,-2.5);
\draw [gray!70] (-2.165,-1.25) ++(-5,5) -- ++(0,-10) -- ++(10,0) -- ++(0,10) -- ++(-10,0) -- ++(4.33,2.5) -- ++(10,0) -- ++(0,-10) -- ++(-4.33,-2.5) -- ++(0,10) -- ++(4.33,2.5);
\draw [ultra thick] (-5,0) -- (5,0)
					  (0,-5) -- (0,5)
					  (-2.165,-1.25) -- (2.165,1.25);
\filldraw [black] (0,0) circle (3pt)
					(-5,0) circle (3pt)
					(5,0) circle (3pt)
					(0,-5) circle (3pt)
					(0,5) circle (3pt)
					(-2.165,-1.25) circle (3pt)
					(2.165,1.25) circle (3pt);
\end{tikzpicture}
\caption{The $3d$ singular locus $S$: lines propagate from the singular points in the $2d$ faces of $\B$. 
The orthonormal axes correspond to adapted coordinates \cite{coordinates}.}
\label{fig:3d}
\end{figure}
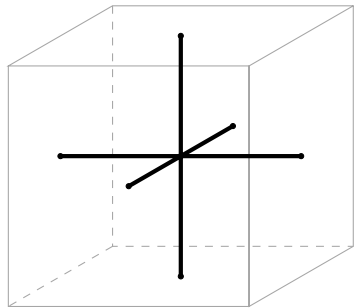

As already stated, when $d=2$ or $d=3$ a non-zero Chern number prevents the existence of a continuous 
and periodic Bloch frame; however, they can be chosen to be smooth, or even analytic, when $d=1$ \cite{Kohn59}. 
Going to polar coordinates, this allows to construct Bloch frames which are smooth along the radial coordinate in $\B$; 
a loss of regularity is necessary in the angular direction. Singularities are then concentrated at a point (when $d=2$, 
cf. \cite{ThonauserVanderbilt2006})  or along lines (when $d=3$), near which the growth of the gradient of the Bloch frame is inversely proportional to the distance to the singular locus $S$, 
namely $|\nabla v_{a, \bk}| \asymp |\dist(\bk, S)|^{-1}$.  The typical singular locus for a $3$-dimensional Brillouin zone is 
shown in Figure~\ref{fig:3d}.  A detailed analysis \cite{companion} shows that this estimate implies \eqref{X^2s_control} for the associated Wannier functions, yielding our first new result. 


We now prove the second novel result of this paper, namely item \ref{item:non-trivial}: we consider a system for which 
\eqref{Finite X^2} holds, so that \eqref{eqn:Dirichlet energy} is also true, and we prove that the manifold of occupied states is topologically trivial by showing that its first Cher numbers are zero ($d \leq 3$). 
We use the so-called \emph{abelian Berry connection}  {\cite{coordinates}}
\begin{equation} \label{A_Berry}
\A = - \iu \sum_{j=1}^{d} \sum_{a=1}^{m} \scal{v_{a,\bk}}{\partial_ j v_{a,\bk}} \di k_j
\end{equation}
where $\partial_j$ means $\frac{\partial}{\partial{k_j}}$. 
A straightforward computation, 
using only the Leibnitz property and $P(\bk) = \sum_a \ket{v_{a, \bk}}\bra{v_{a,\bk}}$, 
shows that its curvature, locally expressed as  $\Omega = \di \A $, is given by 
\begin{align} \label{Curvature}
\Omega = &  \sum_{i < j}   \sum_{a=1}^m  2 \, \mathrm{Im} \scal{\partial_i v_{a,\bk}}{\partial_j v_{a,\bk}}  \di k_i \wedge \di k_j  \\
              = &   - \iu \sum_{i < j} \Tr \Big( P(k) \left[ \partial_i P(k), \partial_j P(k) \right] \Big) \, \di k_i \wedge \di k_j
              \nonumber 
\end{align}
and is thus manifestly gauge invariant. Here, $\Tr$ is the trace in the Hilbert space in which $P(\bk)$ acts, 
which depends on the model considered.  

For simplicity, we assume hereafter that all singularities have the form 
\begin{equation} \label{eqn:Singularity}
|\nabla v_{a, \bk}| \asymp |\dist(\bk, S)|^{- \beta}
\end{equation}
for some $\beta >0$. We emphasize that this assumption is actually not necessary \cite{companion}. 

For $d=2$, topology is characterized by a unique Chern number,  which is proportional to the integral of $\Omega$ over the whole Brillouin torus, namely $c_1(P) = \frac{1}{2\pi} \int_{\B} \Omega \in \Z$. 
Under the simplifying assumption \eqref{eqn:Singularity}, we can use the argument in 
\cite{ThonauserVanderbilt2006}: a small disk $B_{\delta}$ of radius $\delta >0$ is removed from the (centered) 
Brillouin zone.  Hence, for any periodic Bloch frame $v_{\bk}$ one has 
$$
c_1(P) 
 = \frac{1}{2\pi} \int_{\B} \di \A  
= \frac{1}{2\pi} \lim_{\delta \to 0} \left(  \int_{\partial \B} \A  - \int_{\partial B_{\delta}} \A \right)
$$ 
where $\partial \B$ is the boundary of the Brillouine zone. The first integral vanishes since $v_{\bk}$ 
is periodic. The second integral depends on the singularity: in the limit $\delta \to 0$,  it yields zero if 
$\beta < 1$, a finite number if  $\beta =1$, and $+ \infty$ if $\beta>1$. Hence, if  $c_1(P) \neq 0$, one concludes 
that $\beta =1$, which is however incompatible with the assumption \eqref{eqn:Dirichlet energy}. 
Therefore, $c_1(P)=0$ must hold true, and topological triviality is proved.   

The case $d=3$ goes along the same lines, as detailed in 
\cite{companion}. For $d=3$ the topological triviality is characterized by three Chern numbers \cite{Panati2007}, 
which correspond to the integral of $\Omega$ over the $2$-dimensional tori $\B_{i,j} \subset \B$, where \eg $\B_{1,2} = \set{\bk \in \B: k_3= 0}$ and so on. By removing suitable disks as above, 
one concludes that if
$\int_{\B_{1,2}} \Omega \neq 0 $  a typical singularity in the form \eqref{eqn:Singularity} with $\beta =1$ appears. 
By choosing the $k_3$-axis tangent to the singular locus and using polar coordinates, one has   
$
|\nabla v_{a, \bk}| \asymp |k \sin\theta_{\bk}|^{-1}    
$
which is however not compatible with \eqref{eqn:Dirichlet energy}. Hence, $\int_{\B_{i,j}} \Omega$
 must vanish for every $i,j$ with $i<j$. We conclude that, if there exists a system of CWFs such that \eqref{Finite X^2} holds true, 
then all the Chern numbers vanish 
and the system is topologically trivial also for $d \leq 3$. 



In this Letter,  we considered a gapped periodic quantum system, described by a family of orthogonal projectors 
$\set{P(\bk)}$ satisfying  property \eqref{eqn:proj_smooth} and either \eqref{eqn:tau-cov} or  \eqref{eqn:P_periodic}. 
This formalism encompasses both tight-binding and continuous models, since it relies only on the gap condition and the periodicity of the system.  Moreover, our approach applies to gases of interacting electrons modeled within the HF approximation, provided one identifies $P(\bk)$ with $\rho(\bk)$, as in the decomposition \eqref{rho decomposition} of the $1$-body density matrix.  

We showed that whenever the second moment of the position operator $\langle X^2 \rangle_{w}$ is finite, for an arbitrary system $w$ of CWFs, then all the Chern numbers must be zero. Since the latter are related to the transverse conductivity of a $2$-dimensional electron gas \cite{TKNN82, Haldane88}, we conclude that   $\langle X^2 \rangle_{w} < + \infty$ implies the vanishing of the Hall conductivity.  Viceversa, whenever the system is {in a topologically non-trivial phase} (as it may happen when time-reversal symmetry is broken), the system state is delocalized, in the sense that $\langle X^2 \rangle_{w} = + \infty$ for every possible choice of CWFs.  

The latter results have crucial implications for many realms of condensed matter physics. 
For example, recent progresses in flat band superconductivity \cite{PeottaTorma2015}
show that for  $2d$ systems the superfluid weight $D_{\rm s}$ 
satisfies  $D_{\rm s} \geq |c_1(P)|$, where $c_1(P)$ is the first Chern number and $P$ is the relevant family of projectors. 
In view of our localization dichotomy, a non-vanishing Chern number implies the delocalization of composite Wannier functions, which might be related to the existence of a long-range order associated to the phase transition. 
Similarly, our approach can be applied to cold atoms in optical lattices 
\cite{Trombettoni et al experimental 2005, Jotzu et al 2014}. 
Our result shows that, whenever time-reversal symmetry is  broken, the usual derivation of effective tight-binding models for the dynamics of the condensate \cite{SmerziTrombettoni2003} should be handled with care, 
since in a topologically non-trivial phase the well-localized Wannier functions which are used in 
in the derivation do not exist. These and other consequences will be further investigated in the future. 


%




\begin{thebibliography}{99}
\bibitem{coordinates}
We use adapted coordinates, so that $\bk = \sum_{j=1}^d k_j \, \mathbf{b}_j$ where the $\mathbf{b}_j$ form a basis for the reciprocal Bravais lattice.
\end{thebibliography}
\end{document}